\documentstyle[12pt,procsla]{article}
  
  \catcode`\@=11
  \long\def\@makefntext#1{
  \protect\noindent \hbox to 3.2pt {\hskip-.9pt  
  $^{{\ninerm\@thefnmark}}$\hfil}#1\hfill}		
  
  \def\@makefnmark{\hbox to 0pt{$^{\@thefnmark}$\hss}}  
  	
  \def\ps@myheadings{\let\@mkboth\@gobbletwo
  \def\@oddhead{\hbox{}
  \rightmark\hfil\ninerm\thepage}   
  \def\@oddfoot{}\def\@evenhead{\ninerm\thepage\hfil
  \leftmark\hbox{}}\def\@evenfoot{}
  \def\sectionmark##1{}\def\subsectionmark##1{}}
  
\begin{document}
\rightline{{\tt UM-P-99/08,\ RCHEP 99/04}}
\vspace{5mm}
  
  \centerline{\normalsize\bf NEUTRINO PHYSICS AND THE MIRROR WORLD\footnote{Talk at 
{\it 8th International Symposium on Neutrino Telescopes}, Venice, February 1999.}}
  \baselineskip=16pt
  
  \centerline{\footnotesize R. R. VOLKAS}
  \baselineskip=13pt
  \centerline{\footnotesize\it School of Physics, Research Centre for High Energy Physics,
The University of Melbourne}
  \baselineskip=12pt
  \centerline{\footnotesize\it Parkville 3052, Australia}
  \centerline{\footnotesize E-mail: r.volkas@physics.unimelb.edu.au}

  \vspace*{0.9cm}
  \abstracts{Improper Lorentz Transformations can be retained as exact symmetries of Nature 
if the particle content and gauge group of the Standard Model are doubled. The resulting
``Exact Parity Model (EPM)'' sees each ordinary
particle paired with a mirror analogue. If neutrinos have mass and if they mix,
then the EPM predicts that each ordinary neutrino will be maximally mixed with its mirror
neutrino partner. This provides a very simple explanation for the very large mixing angle
observed for atmospheric muon neutrinos by SuperKamiokande and other experiments. Maximal
mixing for electron neutrinos is also well motivated by the solar neutrino problem. If
small interfamily mixing is switched on, then the LSND anomaly can also be accomodated by
the EPM. The EPM thus provides a unified, simple and to some extent predictive framework
for explaining all of the anomalous neutrino data. This talk will briefly review the EPM or
mirror neutrino scenario.}
   
  \normalsize\baselineskip=15pt
  \setcounter{footnote}{0}
  \renewcommand{\thefootnote}{\alph{footnote}}
  \section{Full Lorentz Invariance: Proper and Improper}
  
For many decades there was a strong theoretical and/or aesthetic prejudice for fundamental
physical laws that were symmetric under both spatial and temporal reflection (parity and
time-reversal invariance). When the $V - A$ character of weak interactions was established
in the late 1950's, exact parity invariance was apparently empirically
falsified.\cite{wu,ly} Soon
after, the discovery of $CP$ violation apparently falsified exact time-reversal
invariance.\cite{cf}
It seemed as if only the Proper Lorentz Transformations were true symmetries of Nature.

However, Lee and Yang recognised from the outset that parity can be retained as an exact
symmetry, despite the $V - A$ character of weak interactions, provided that the ordinary
particle spectrum is doubled.\cite{ly} From a modern theoretical and phenomenological
perspective
this requires every ordinary lepton, quark, gauge boson and Higgs boson to be paired with a
mirror analogue.\cite{flv1} Since the ordinary and mirror particle sectors are but weakly
coupled to
each other, the resulting scenario is phenomenologically viable. The violation of parity
invariance
is therefore, remarkably, still an open question. A simple argument to be presented below
shows that if the above type of exact parity symmetry exists in Nature, then so necessarily
does a form of time reversal invariance. So, it is possible for the full Lorentz Group to be
a completely unbroken symmetry of Nature!\cite{flv1}

Of particular interest is the fact that the observed atmospheric and solar neutrino
anomalies may be the first experimental manifestation of the mirror matter sector
(mirror neutrinos to be specific).\cite{flv2,f,fv} In order to further test this proposal,
neutral current
based measurements probing both the atmospheric and solar anomalies are vital. Such
measurements will determine whether the relevant ordinary neutrinos are transforming into
other ordinary neutrinos, or into something more exotic such as mirror or sterile
neutrinos.

The gauge theoretic construction of a theory with exact parity symmetry is very easy to
understand.\cite{flv1,fv} Consider a theory defined by a parity {\it violating} Lagrangian
${\cal L}$
which has gauge group $G$. This theory may, for instance, be the minimal Standard Model, or,
more pertinently, the Standard Model augmented by nonzero neutrino masses and mixings. For
every field $\psi$ in ${\cal L}$ introduce a mirror or parity partner $\psi'$. For
spin-$1/2$ fields this requires of course that the $\psi'$ have opposite chirality to
the $\psi$. The
fields $\psi'$ are singlets under $G$ but transform under a gauge group $G'$ which is
isomorphic to $G$, while the $\psi$'s are correspondingly required to be singlets
under $G'$.
The fields $\psi$ and $\psi'$ are placed into identical multiplets under their
respective gauge groups $G$ and $G'$, and the discrete parity symmetry (schematically $\psi
\leftrightarrow \psi'$) is enforced. The resulting Lagrangian is
\begin{equation}
{\cal L}_{\rm total}(\psi,\psi') = {\cal L}(\psi) + {\cal L}'(\psi') + {\cal
L}_{\rm int}(\psi,\psi'),
\end{equation}
where ${\cal L}'$ is exactly the same function of the $\psi'$ fields as ${\cal L}$ is of the
$\psi$ fields.\footnote{The dependence of the Lagrangian on first derivatives of the fields
is of course understood.}
The extremely important interaction term ${\cal L}_{\rm int}$ describes any
gauge and parity invariant renormalisable coupling terms between the ordinary and mirror
sectors. The above procedure was first carried out for the minimal Standard Model in
the first paper quoted under Ref.\cite{flv1},
where it was shown that parity was a symmetry of the vacuum as well as the Lagrangian for a
large region in Higgs potential parameter space. We will focus on this parameter space
region
from now on.\footnote{By extending the Higgs sector, it is possible to spontaneously break
the parity symmetry together with the electroweak and mirror-electroweak gauge
symmetries.\cite{bm}}
We call the resulting theory the ``Exact Parity Model (EPM)''.

The ordinary and mirror sectors are coupled by gravitation and ${\cal L}_{\rm int}$. The
gravitational coupling is very interesting from the point of view of cosmology and the dark
matter problem,\cite{bk} but will not be further discussed here. The nongravitational
effects in
${\cal L}_{\rm int}$ in general feature photon -- mirror photon, $Z$ -- mirror $Z$ and Higgs
-- mirror Higgs mixing. These particles are singled out because they are neutral under the
electromagnetic and colour forces and their mirror analogues, so there are no exact
conservation laws to prevent mixing. (Electron -- mirror electron mixing is forbidden by
both ordinary and mirror electric charge conservation, for instance.) Unfortunately,
cosmological constraints from Big Bang Nucleosynthesis make it unlikely that these effects
will be seen in the laboratory.\cite{cg}

We now come to the crux of the matter: {\it since neutrinos and mirror neutrinos are
electrically neutral and colourless, they will in general mix if they also have nonzero
masses.} Furthermore, we will see in the next section that the exact parity symmetry forces
the mixing angle between an ordinary neutrino and its mirror partner to be the maximal value
of $\pi/4$.

I close this section with two brief comments. (i) Let the exact parity symmetry be denoted
by $P'$. Note that it is different from the usual (broken) parity symmetry $P$.\footnote{For
instance, the left-handed electron is transformed into the right-handed mirror-electron by
$P'$, whereas it is transformed into the right-handed electron by $P$.} However, standard
$CPT$ is still of course an exact symmetry of the theory. We can therefore define a
non-standard time-reversal invariance $T'$ through $CPT = P'T'$ that must necessarily be
exact if $P'$ is exact. The full Lorentz Group, including all Improper Transformations, is
thus a symmetry of the theory.\cite{flv1} (ii) It is amusing to compare the `exact parity'
idea with
spacetime supersymmetry. Both extend the Proper Lorentz or Poincar\'e Group, and both
require degree-of-freedom doubling. A crucial difference, though, is that phenomenology
forces spacetime supersymmetry to be broken. The resulting proliferation of
soft-supersymmetry breaking parameters has no analogue in the EPM.

\section{Phenomenology of Mirror Neutrinos}

Under the exact parity symmetry $P'$, an ordinary neutrino field $\nu_{\alpha}$ (where
$\alpha = e, \mu, \tau$) transforms into its mirror partner field $\nu'_{\alpha}$ as per
\begin{equation}
\nu_{\alpha L} \to \gamma_0 \nu'_{\alpha R}.
\end{equation}
From basic quantum mechanics, we know that the exact $P'$ symmetry forces the parity
eigenstates to also be mass eigenstates. In the absence of interfamily mixing, this means
that the two mass eigenstates $\nu_{\alpha \pm}$ per family take the form
\begin{equation}
| \nu_{\alpha \pm} \rangle = \frac{1}{\sqrt{2}} \left( |\nu_{\alpha}\rangle \pm
|\nu'_{\alpha}\rangle \right).
\end{equation}
The positive and negative parity states, $\nu_{\alpha +}$ and $\nu_{\alpha -}$
respectively, in general have arbitrary masses. The oscillation parameter 
$\Delta m^2_{\alpha+ \alpha-} \equiv |m^2_{\nu_{\alpha +}} - m^2_{\nu_{\alpha -}}|$ is
therefore
free. The mixing angle, however, is forced by $P'$
symmetry to have the maximal value of $\pi/4$.\cite{flv2,f,fv}

Interestingly, SuperKamiokande and other experiments have observed a disappearence of
atmospheric muon-neutrinos in a manner which favours maximal mixing with another flavour
$\nu_x$.\cite{sktalk} Current results preclude $x = e$, but allow both $x = \tau$ and $x =
s$\cite{all} (where $s$
stands for ``sterile''). It is natural in the EPM to identify $\nu_x$ with the effectively
sterile mirror muon-neutrino $\nu'_{\mu}$.\cite{f,fv} The maximal mixing angle for the
$\nu_{\mu} -
\nu'_{\mu}$ subsystem is a simple and characteristic prediction of the EPM that is strongly
supported by experiment.\footnote{For attempts to explain the large mixing angle in the case
of $\nu_x$ identified as $\nu_{\tau}$ see, for instance, Ref.\cite{a}.} The $\Delta
m^2_{\mu+
\mu-}$ oscillation parameter is adjusted to agree with the measurements. This requires it to
be in the $10^{-3} - 10^{-2}$ eV$^2$ range.\cite{sktalk,fvy1}

The experimental discrimination between $\nu_x = \nu_s$ and $\nu_x = \nu_{\tau}$ is a vital
further test of this proposal. Hope for progress in this area in the immediate future lies
with SuperKamiokande atmospheric neutrino data and the K2K long baseline
experiment.\cite{s} The
basic requirement is a neutral current measurement, since $\nu_\tau$ is sensitive to this
interaction while $\nu_s$ and $\nu'_\mu$ are not.  SuperKamiokande has quoted a measured
value for the atmospheric neutrino induced $\pi^0/e$ ratio (see Ref.\cite{sktalk} for the
present
status) that cannot discriminate between the two possibilities because of a large
theoretical uncertainty in the $\pi^0$ production cross-section.  The measurement of this
cross-section by the K2K long baseline experiment is thus of great importance. This may
allow a discrimination based on a zenith-angle averaged atmospheric neutrino induced
$\pi^0/e$ ratio by SuperKamiokande within about a year from the time of writing.\cite{l} 
It
should be noted, however, that the $\nu_\mu \to \nu'_\mu\, (\nu_s)$ case predicts that the
actual $\pi^0/e$ ratio will be about $0.8$ times the no-oscillation or $\nu_{\mu} \to
\nu_\tau$ expectation.\cite{fpriv} If the SuperKamiokande central value were to be around
$0.9$, then
the remaining systematic error would still be too large to discriminate between the
possibilities. A cleaner discrimination, which however requires significantly greater
statistics, lies in the future through the $\pi^0$ up-down asymmetry.\cite{dg} The K2K
experiment
could in principle discriminate between the two possibilities on its own by comparing the
neutral- to charged-current rates at the near and far detectors. However, inadequate
statistics at the far detector (SuperKamiokande) may preclude a useful result. However,
provided $\Delta m^2_{\mu+\mu-}$ is sufficiently large, it should at the very least confirm
$\nu_\mu$ disappearence. Looking slightly further into the future, the long baseline
experiment MINOS and the proposed CERN--Gran-Sasso long baseline experiments should provide
important information.\cite{lbl} 

The solar neutrino anomaly provides further motivation for the maximal mixing feature of the
EPM.\cite{flv2,fv} Consider the maximally mixed $\nu_e - \nu'_e$ subsytem in the zero
interfamily mixing
limit. For the $10^{-3} \stackrel{<}{\sim} \Delta m^2_{e+e-}/{\rm eV}^2 \stackrel{<}{\sim}
10^{-10}$ range, the maximal $\nu_e \to \nu'_e$ oscillations are consistent with
disappearence experiments and lead to an energy-independent day-time solar neutrino flux
reduction by $50\%$. This is consistent with four out of the five solar event rate
meansurements relative to the latest standard solar model calculations.\cite{solar} (The
Chlorine
experiment sees a greater than $50\%$ deficit.) Since this talk was given, Guth et
al.\cite{g}\ have
emphasised that the night-time oscillation-affected solar neutrino rate differs from the
day-time rate, even if the vacuum mixing angle is maximal. This leads to some
energy-dependence in the night-time flux suppression, and provides an interesting further
test. Preliminary calculations show that the day-night asymmetry for the $\nu_e \to \nu'_e$
case (or the $\nu_e \to \nu_s$ case with maximal mixing) is potentially observable for the
range $6 \times 10^{-8} \stackrel{<}{\sim} \Delta m^2_{e+e-}/{\rm eV}^2 \stackrel{<}{\sim} 2
\times 10^{-5}$, with the range $2 \times 10^{-7} \stackrel{<}{\sim} \Delta
m^2_{e+e-}/{\rm eV}^2 \stackrel{<}{\sim} 8 \times 10^{-6}$ already disfavoured by
the data.\cite{cfv} 

The future KAMLAND experiment will probe the $10^{-3} \stackrel{<}{\sim} \Delta
m^2_{e+e-}/{\rm eV}^2 \stackrel{<}{\sim} {\rm few}\ \times 10^{-5}$ regime by looking for
$\overline \nu_e$ disappearence.\cite{kamland}\footnote{The $\nu_e \to \nu'_e$ mode also has
potentially
observable consequence for atmospheric $\nu_e$'s for this parameter range.\cite{bfv}} 
Another extremely important future test is the neutral to
charged current event rate ratio that will be measured by SNO.\cite{sno} The mirror
electron-neutrino
$\nu'_e$ is effectively a sterile flavour, so SNO should measure the ``standard'' value for
this ratio.

If $\Delta m^2_{e+e-}/{\rm eV}^2$ is in the $10^{-10} - 10^{-11}$ range, then ``just-so''
oscillations result.\cite{justso} One amusing possibility\cite{cfv} is the following: as
$\Delta m^2_{e+e-}$ is
reduced from the range considered in the previous paragraph into the just-so regime, the
energy at which the averaged oscillations give way to coherent just-so behaviour decreases.
For some value, this transition will happen within the energy range probed by
SuperKamiokande. This could possibly be the origin of the mysterious high-energy spectral 
feature reported by SuperKamiokande!\cite{solar} (This type of idea was first examined in
the context
of $\nu_e$ oscillations into an active flavour in Ref.\cite{bfl}.)

So, putting the above in a nutshell, we have KAMLAND probing the high range for $\Delta
m^2_{e+e-}$, the day-night asymmetry being used in the intermediate range, and just-so
signatures such as seasonal variation and Boron neutrino energy spectrum distortion probing
the low $\Delta m^2_{e+e-}$ regime. The range between about $6 \times 10^{-8}$ eV$^2$ and
the beginning of the just-so region appears to have no characteristic signature other than
the $50\%$ energy independent flux suppression. Furthermore, the crucial neutral current
measurement at SNO will test the general idea that solar neutrinos are disappearing into
sterile states of some sort for the whole $\Delta m^2_{e+e-}$ range of interest.

The above analysis saw interfamily mixing set to zero. Certainly, small interfamily
neutrino mixing is well motivated by the small mixing observed for
the quark sector. However, it is unlikely that this mixing exactly vanishes. I will
now comment on three possible consequences of interfamily mixing.

First, the LSND anomaly\cite{lsnd} can be trivially accomodated within the EPM by switching
on $\nu_e -
\nu_{\mu}$ mixing with the appropriate parameter choices.\cite{fv} The LSND parameter regime
does not
significantly modify the solar and atmospheric neutrino scenario discussed above.

Second, the solar neutrino flux depletion can be due to an amalgam of vacuum $\nu_e \to
\nu'_e$ oscillations and MSW interfamily transitions.\cite{vw} This leads to characteristic
energy-dependent flux depletions depending on the precise oscillation parameter range
chosen. Further, the neutral to charged current induced event rate ratio to be measured by
SNO can take on values intermediate between the extreme cases of $\nu_e \to \nu_{\rm
active}$ only and $\nu_e \to \nu_s$ only.

Third, it turns out that small interfamily mixing is well motivated from cosmology, a topic
I very briefly review in the next section.

\section{Cosmology}

The tale of how neutrino oscillations affect early universe cosmology is long and
complicated. I will pass over it lightly here, just for the sake of completeness, without
much in the way of explanations. Please consult, for example,
Refs.\cite{cosmo1,cosmo2,cosmo3,epm} for further details.

Cosmology and ordinary-mirror (and ordinary-sterile) neutrino oscillations present
challenges to each other. On the one hand, it had long been thought that sterile neutrinos
ought to mix but weakly with ordinary neutrinos lest the reasonably successful Big Bang
Nucleosynthesis (BBN) predictions be spoiled. In particular, it was thought that a
$\nu_{\mu} \to \nu_s$ solution to the atmospheric neutrino problem would have necessarily
implied the thermal equilibration of the sterile flavour prior to the BBN epoch, and thus
would have increased the expansion rate of the universe. Recall that the expansion rate of
the universe during BBN is driven by the relativistic degrees of freedom in the plasma, with
``neutrino flavour number $N_{\nu}$'' being a convenient measure. In the minimal Standard
Model $N_{\nu} = 3$, while one thermally equilibrated sterile flavour in addition to the
ordinary neutrinos produces $N_\nu = 4$. There is some confusion in interpreting primordial
element abundance data at present, but it is arguable that $N_\nu < 4$ is
preferred.\cite{Nnu} So, it had been thought that a large region of active-sterile
oscillation parameter space was at least disfavoured by BBN. This problem was seen to be
much more acute for the EPM than for models with a single extra sterile state, because of
the {\it three} mirror neutrino flavours as well as the mirror photons, electrons and
positrons. Prior wisdom would have concluded that the EPM ruined BBN and was therefore
unlikely to be true. Thus cosmology challenged sterile and mirror neutrino models.

On the other hand, the discovery of relic neutrino asymmetry amplification, driven by the
ordinary-mirror or ordinary-sterile neutrino transitions themselves, showed that the
previous pessimism was misplaced: a very natural mechanism for reconciling BBN with sterile
or mirror neutrinos, born out of the apparently problematic neutrino scenario itself,
actually existed all along but had been missed.\cite{cosmo2} The basic point is that large
relic neutrino
asymmetries (neutrino chemical potentials) will, in a certain large region of oscillation
parameter space, be inevitably created via a positive feedback process from the tiny
$CP$ asymmetry (baryon asymmetry for instance) of the high-temperature background plasma.
The large matter (Wolfenstein) effective potentials so induced then damp further
ordinary-mirror or ordinary-sterile transitions and lead to quite acceptable BBN
predictions (for the appropriate region of parameter space).

The full story of BBN in the presence of ordinary-mirror/sterile transitions is complicated
because many different oscillation modes are in general involved. We have discussed above
how the excitation of mirror or sterile neutrinos prior to BBN increases the expansion rate
as quantified through $N_\nu$. But there is another important effect: a fairly large
electron neutrino asymmetry will be created before and during BBN given appropriate
oscillation parameters. This asymmetry will directly affect BBN reaction rates and will
alter the primordial Helium abundance so as to mimic {\it either} a negative {\it or} a
positive contribution to the effective neutrino number during BBN. A detailed numerical
calculation is often necessary to determine the final BBN outcome. Such calculations have
been performed for a couple of models featuring a single sterile flavour.\cite{cosmo3} They
have demonstrated that strong ordinary-sterile neutrino mixing can be reconciled with BBN
for realistic sterile neutrino models via the interesting physics just discussed.

The first full analysis of neutrino asymmetry evolution and BBN in the Exact Parity Model
was completed after this Symposium.\cite{epm} It demonstrated that the EPM scenario outlined
above is consistent with primordial element abundance measurements for a large region of
oscillation parameter space. It turns out that this parameter space region requires some
small interfamily mixing.

The challenge for observational cosmology, then, is to pin down cosmological
parameters precisely enough to test early universe neutrino physics in some detail.
Continuing primordial element abundance measurements will help, but much dramatic new
information is likely from the cosmic microwave background anistropy measurements promised
by the future MAP and PLANCK satellite missions.\cite{kk}

\section{Conclusion}

The Exact Parity Model predicts that if ordinary neutrinos mix with their mirror partners,
then they mix maximally. This has been proposed as a very natural and simple explanation of
the very large mixing angle deduced from atmospheric $\nu_{\mu}$ disappearence measurements.
In addition, maximal oscillations of the $\nu_e$ into its mirror partner are well motivated
by most of the solar neutrino data. With small interfamily mixing switched on, the LSND
anomaly can be explained by ordinary $\nu_e \to \nu_\mu$ oscillations. The EPM offers a
theoretically elegant solution to all of the neutrino puzzles within a model that had as its
original motivation the retention of the full Lorentz Group as an exact symmetry of Nature.
The model also has some very interesting consequences for early universe cosmology,
particularly the process of Big Bang Nucleosynthesis. In addition to the important results
that continue to be produced by SuperKamiokande, we await with interest upcoming experiments
-- such as K2K, SNO, KAMLAND and others -- that will provide further crucial tests of the
Exact Parity Model.

\section{Acknowledgements}

I would very much like to thank Professor Milla Baldo Ceolin for organising this stimulating
symposium in such an inspirational setting. I would also like to thank the participants of
the symposium for interesting discussions and for their informative presentations. I warmly
acknowledge my very wise long time collaborator Robert Foot, and my present students Nicole
Bell, Roland Crocker and Yvonne Wong for many fruitful scientific discussions.



  \end{document}